\documentstyle[12pt,epsfig]{article}
\renewcommand{\bar}[1]{\overline{#1}}

\begin{document}
\bigskip\bigskip
\begin{flushright}
GEF-Th-2/2008\\
PCCF RI 0802\\
May 2008
\end{flushright}

\begin{center}
{\large \bf Model Independent Tests \\
for Time Reversal and CP Violations \\
and for CPT Theorem \\
in ${\bf\Lambda_b}$ and ${\bf{\bar\Lambda}_b}$ Two Body Decays}
\end{center}

\vspace{12pt}

\begin{center}
 {\bf E.~Di~Salvo\footnote{Elvio.Disalvo@ge.infn.it}\\}

 {Dipartimento di Fisica and I.N.F.N. - Sez. Genova, Via Dodecaneso, 33 \\-
 16146 Genova, Italy\\}
 
 {\bf Z.~J.~Ajaltouni\footnote{ziad@clermont.in2p3.fr}\\}

 {Laboratoire de Physique Corpusculaire de Clermont-Ferrand, \\
IN2P3/CNRS Universit\'e Blaise Pascal, 
F-63177 Aubi\`ere Cedex, France\\}
\end{center}

\vspace{10pt}
\begin{center} {\large \bf Abstract}

Weak decays of beauty baryons like $\Lambda_b ({\bar{\Lambda}_b})$ into $\Lambda 
({\bar{\Lambda}})$ and $V(J^P=1^-)$, where both decay products are polarized, offer 
interesting opportunities to perform tests of time reversal and CP violations and of 
CPT invariance. We propose a model independent parametrization, via spin density 
matrix, of the angular distribution, of the polarizations and of some polarization 
correlations of the decay products. The transverse component of the polarization and 
two polarization correlations are sensitive to time reversal violations. Moreover 
several CP- and CPT-odd observables are singled out.

\end{center}

\vspace{10pt}

\centerline{PACS Numbers: 11.30.Er, 12.39.-x, 12.39.ki, 13.30.-a, 14.20.Mr}

\newpage

\section{Introduction}

The interest in CP violations (CPV) and time reversal violations (TRV) has been 
increasing
in last years[1-10]. The reason is that, although some CPV and also a direct 
TRV\cite{cpl} have been detected experimentally, the nature of such symmetry 
violations has not yet been clarified. More precisely, the prediction of the size of 
the violation in some weak decays is strongly model dependent, which stimulates 
people to search for signals of new physics\cite{alv,cgn,bld,bld2,ibl,chp,utf} (NP), 
beyond the standard model (SM). For example, the decays involving the transition  
\begin{equation}
b \rightarrow s \label{trans}
\end{equation}
present CPV parameters, like the $B^0-{\bar B}^0$ mixing phase\cite{ibl,utf} and the 
transverse polarization of spinning decay products of $\Lambda_b$\cite{alv}, which 
are very small in SM predictions, but are considerably enhanced in other models. In 
particular, recent signals of NP have been claimed in B decays: the CP violating 
phases of $B\rightarrow\pi K$\cite{ibl} and $B_s\rightarrow\Phi J/\Psi$\cite{utf} may 
be considerably greater than predicted by SM. Also $\Lambda_b$ 
decays\cite{alv,cg,blss,aj} are suggested as new sources of CPV and TRV parameters, 
especially in view of the abundant production of this resonance in the forthcoming 
LHC accelerator. 

As regards direct TRV, only one evidence\cite{cpl,ap,ak,ge,fg,elm} has been given so 
far, and assuming the Bell-Steinberger\cite{bes} relation, which might be violated to 
few percent\cite{bs}. Lastly the CPT theorem, valid for local field theories, has 
been tested to a great precision in the neutral kaon decay\cite{pdg}, but not in 
other situations: for example, it has never been checked in decays involving the 
$b$-quark, furthermore a meaningful size of uncertainty remains in 
$K^{\pm}\rightarrow \pi^{\pm} \pi^0$\cite{bs}. 
    	
The aim of the present paper is to suggest model independent tests of TRV, CPV and 
CPT invariance in hadronic $\Lambda_b$ and $\bar{\Lambda}_b$ decays of the type
\begin{eqnarray}
\Lambda_b (\bar{\Lambda}_b) &\rightarrow& \Lambda (\bar{\Lambda}) V, \label{decl}
\end{eqnarray}
$V$ denoting a $J^P = 1^-$ resonance, either the $J/\psi$ or a light vector meson, 
like $\rho^0, \omega$. Each resonance decays, in turn, to more stable particles, 
like, {\it e. g.}, $\Lambda\rightarrow p\pi^-$, $J/\psi\to\mu^+\mu^-$, so that one 
has to consider a typical {\it cascade decay}. 
 A previous paper\cite{aj} had been devoted to the subject. Now we parametrize, by 
means of the spin density matrix (SDM), the angular distribution and the 
polarizations of the decay products, without introducing any dynamic assumption at 
all. Then we study the behavior of these observables under CP and T, singling out 
those which are sensitive to T, CP and CPT violations. Our approach resembles the one 
proposed by Lee and Yang\cite{ly} and by Gatto\cite{ga} many years ago, to use 
hyperon decays for the same tests.
However, as we shall see, a hadronic two-body weak decay involving two spinning 
particles in the final state - never proposed before - presents some advantages over 
hyperon decays\cite{ly,ga,chp}, where one of the two final particles is spinless.    

In sect. 2 we derive the expressions of the spin density matrices, angular 
distribution and polarizations of the decay products in the above mentioned decays. 
In sect. 3 we present a parametrization of the angular distribution and of 
polarizations. In sect. 4 we suggest tests for TRV, CPV and CPT. Lastly we conclude 
with some remarks in sect. 5.  
 
\section{Angular Distribution and Polarization Vectors of the Decay Products} 

In order to deal with the angular distribution and the polarization of the 
intermediate resonances, $ \Lambda$ and $V$, coming from $\Lambda_b$ decay, the best 
suited method consists of applying the relativistic helicity formalism
pioneered by Jacob and Wick\cite{jw} and reformulated later by Jackson\cite{jk} (see 
also\cite{bls,chu}). This formalism presents some advantages: 

~(i) thanks to its definition, $ \lambda = {\vec j} \cdot {\hat p}, \ \ {\mathrm 
where} \ \  \vec j = \vec {\ell} + \vec s
 \ {\mathrm and} \ \  \hat p = {\vec p}/|{\vec p}|  \ $, the helicity of a particle 
of spin $\vec s \ {\mathrm and \ momentum} \ 
 \vec p \ $does not depend on its orbital angular momentum $\vec \ell$ and it is 
rotationally invariant; 

(ii) $\lambda$ equals the spin projection along $\vec p$ in the resonance rest frame.  
\newline
These physical properties can be applied just to {\it cascade decays} of the type 
described above, {\it i. e.},
\begin{equation} 
R_0 \to R_1 + R_2,  \  {\mathrm followed \ by }  \  R_1  \to  a_1 + b_1  \  {\mathrm 
and}  \  R_2  \to  a_2 + b_2, \label{dec}
\end{equation} 
provided we take, in the rest frame of the resonance $R_1$ or $R_2$, the quantization 
axis parallel to its momentum in the $R_0$ rest frame.  The helicity of $R_i$ ($i$ = 
1,2), computed in the $R_0$ rest frame, is equal to the projection of its total 
angular momentum along this quantization axis in the $R_i$  rest frame. In our case 
we identify $R_0$ with $\Lambda_b$, $R_1$ with $\Lambda$ and $R_2$ with $V$.

\noindent
In the following, the formalisms of helicity and SDM  will be intensively used by 
specifying different rest frames. 

\subsection{Spin Density Matrices} 

In the standard detector frame the $z$-axis is taken parallel to the incident proton 
beam. For our aims it is more convenient to define a different frame, through the 
three mutually orthogonal unit vectors
$${\vec e}_z \  = \ {\vec n} \ = \ \frac{{\vec p}_p \times {\vec p}_b}{|{{\vec p}_p 
\times {\vec p}_b }|}, ~~~~~~ {\vec e}_x \  =  \ \frac{{\vec p}_p}{|{\vec p}_p|}, 
~~~~~~  {\vec e}_y \ = \ {\vec e}_z  \times {\vec e}_x.$$
Here ${\vec p}_p$ and ${\vec p}_b$ are, respectively, the proton momentum and the 
$\Lambda_b$ momentum. If produced by means of strong interactions - as usually 
assumed for $\Lambda, \Sigma, \Xi, ...$ hyperons -, the $\Lambda_b$ is polarized 
along ${\vec n}$. Therefore
we find it suitable to choose the quantization axis along ${\vec e}_z = {\vec n}$.    

\vskip 0.2cm
\noindent
{\underbar { $\Lambda_b$ SDM }} \\
\noindent
We denote, here and in the following, the $\Lambda_b$ spin by $J$, with $J$ = 1/2. 
Therefore the $\Lambda_b$ SDM reads
\begin{eqnarray}
\rho^{\Lambda_b}  \ = \   {\frac{1}{2}}
(1 + 2{\vec {\cal P}^{\Lambda_b}} \cdot {\vec {\sigma}}).
\end{eqnarray}
Here $\vec \sigma  =  (\sigma_x, \sigma_y, \sigma_z)$, $\sigma_i$ are the Pauli 
matrices and ${\vec {\cal P}^{\Lambda_b}}$ is the polarization vector of $\Lambda_b$.
Defining a $\Lambda_b$ rest frame, whose axes are oriented like those in the 
laboratory frame, the components of ${\vec {\cal P}^{\Lambda_b}}$ result in
\begin{eqnarray}
P^{\Lambda_b}_z \ = \ \frac{1}{2}({\rho}^{\Lambda_b}_{+ +} - {\rho}^{\Lambda_b}_{- 
-}), \ ~~~~ \ P^{\Lambda_b}_x \ = \ 
\Re{({\rho}^{\Lambda_b}_{+ -})}, \ ~~~~ \  P^{\Lambda_b}_y \ = \ 
 -\Im{({\rho}^{\Lambda_b}_{+ -})}.   \label{polr}
\end{eqnarray}
$\rho^{\Lambda_b}_{M M'}$ are the matrix elements of $\rho^{\Lambda_b}$, $M, M' = 
\pm$ denoting the values of the third component of the ${\Lambda_b}$ spin along the 
quantization axis. $\rho^{\Lambda_b}$ verifies the normalization condition 
$Tr({\rho}^{\Lambda_b}) = {\rho}_{+ +} + {\rho}_{- -}  = 1$. The components of the 
polarization vector are regarded as external parameters. Note that, if parity is 
conserved in the production process, we have $P^{\Lambda_b}_x = P^{\Lambda_b}_y = 0.$  

\vskip 0.2cm
\noindent
{\underbar {SDM of the $\Lambda$-V System}} \\
\noindent
The intermediate state in a cascade decay of the type (\ref{dec}) is a composite one, 
consisting of the two spinning particles $\Lambda$ and V. The SDM of this state is 
given by 
\begin{eqnarray}
{\rho}^f  \ = \  {\cal M} {\rho}^{\Lambda_b} {\cal M}^{\dagger}, \label{rof}
\end{eqnarray}
where ${\cal M}$ is the (unitary) operator which describes the decay considered. 
The matrix elements of the SDM ${\rho}^f$ are obtained from (\ref{rof}) by projecting 
the operators involved in that expression onto the initial and final states. 
The latter ones are characterized  by a given three-momentum in the ${\Lambda_b}$ 
center-of-mass system and by a pair of helicities, $\lambda_1$ and $\lambda_2$, 
corresponding to each resonance $R_1$ and$R_2$. Therefore the SDM of this 
two-particle system is endowed with two pairs of indices, {i. e.}, 

\begin {eqnarray} 
{\rho}^f_{\lambda_1  \lambda'_1 \lambda_2  \lambda'_2} &=& {\sum}_{M, M'} 
F^{JM}_{\lambda_1,\lambda_2}(\theta,\phi) {\rho}^{\Lambda_b}_{M M'} 
F^{JM'*}_{\lambda'_1,\lambda'_2}(\theta,\phi), \label{rl00}
\\
F^{JM}_{\lambda_1,\lambda_2}(\theta,\phi) &=& 
<\theta,\phi; \lambda_1,\lambda_2|{\cal M}|JM>. \ ~~~~~~~~ \ \ ~~~~~~~~ \
\end{eqnarray}
\noindent
Here $\theta$  and $\phi$ are, respectively, the polar and azimuthal angle of the 
momentum of the $\Lambda$ resonance in the $\Lambda_b$ rest frame. Note that the 
average value of a given operator ${\cal O}$ over the mixing of states defined by SDM 
(\ref{rl00}) reads
\begin{equation}
\langle {\cal O} \rangle = \sum_{\lambda_1  \lambda'_1 \lambda_2  \lambda'_2} 
{\rho}^f_{\lambda_1  \lambda'_1 \lambda_2  \lambda'_2} 
{\cal O}_{\lambda'_1  \lambda_1 \lambda'_2  \lambda_2}. \label{oper}
\end{equation}
From now on we shall denote this sum over two pairs of indices with the usual symbol 
of "Trace", {\it i. e.}, $Tr({\rho}^f {\cal O})$.

Angular momentum conservation demands 
\begin{equation}
\Lambda'  \ = \  \lambda_1 - \lambda_2 \ ~~~~~~~~~ \ \Lambda'' \ = \ \lambda'_1 
-\lambda'_2, \label{amc}
\end{equation}
where $\Lambda', \Lambda'' = \pm 1/2$.
Therefore the expression (\ref{rl00}) of the SDM can be conveniently transformed to
\begin {eqnarray} 
{\rho}^f_{\lambda_1 \lambda'_1 \Lambda' \Lambda''} \ = \ {\sum}_{M, M'}
F^{JM}_{\lambda_1,\lambda_1 -\Lambda'}(\theta,\phi) {\rho}^{\Lambda_b}_{M 
M'}F^{JM'*}_{\lambda'_1,\lambda'_1 -\Lambda''}(\theta,\phi).\label{rl1}
\end{eqnarray}
The helicity formalism implies
\begin {eqnarray}
F^{JM}_{\lambda_1,\lambda_2}(\theta,\phi) \ = \  N_J  \ A^J_{\lambda_1 , \lambda_2} \ 
D^{J \star}_{M, \Lambda'}{(\phi,\theta,0)}.
\end{eqnarray}
\noindent
Here $N_J = \sqrt{{(2J+1)}/{4 \pi}}$, $D^J_{M, \Lambda'}$ is a rotation matrix 
element and
\begin{equation}
A^J_{\lambda_1 , \lambda_2} \ = \ 4\pi\left(\frac{M_b}{p}\right)^{1/2}
<J,M;\lambda_1 , \lambda_2|{\cal M}|J,M>
\end{equation}
is the rotationally invariant decay amplitude, $M_b$ being the $\Lambda_b$ rest mass 
and $p$ the momentum of $\Lambda$ in the $\Lambda_b$ rest frame.

Now we sum over the indices $ M \ {\mathrm and} \ M'$ in the expression (\ref{rl1}), 
taking into account eqs. (\ref{polr}), and recalling the well-known properties of the 
$D$-functions\cite{chu}. As a result we get
\begin {eqnarray}
{\rho}^f_{\lambda \lambda' \Lambda' \Lambda''} \ &=& \ \frac{1}{4\pi} 
\Big[A_{\lambda,\lambda-\Lambda'} A^*_{\lambda',\lambda'-\Lambda'} 
(1+4\Lambda'P^{\Lambda_b}_1)\delta_{\Lambda',\Lambda''} \nonumber \\
&+& 2A_{\lambda,\lambda-\Lambda'} A^*_{\lambda',\lambda'+\Lambda'}
(P^{\Lambda_b}_2 + 2i \Lambda' P^{\Lambda_b}_3)\delta_{\Lambda',-\Lambda''}\Big]. 
\label{dm11}
\end{eqnarray}
\noindent
Here we have dropped the index $J$ from the $A$-amplitudes and the index 1 from 
helicities. Moreover we have set 
\begin{equation}
P^{\Lambda_b}_1 = {\vec{\cal P}}^{\Lambda_b} \cdot {\hat p}, \ ~~~~~~ \ 
P^{\Lambda_b}_2 = {\vec{\cal P}}^{\Lambda_b} \cdot {\vec e}_N, \ ~~~~~~ \ 
P^{\Lambda_b}_3 = {\vec{\cal P}}^{\Lambda_b} \cdot {\hat r}, \label{scp}
\end{equation}
where ${\hat p}$ is the unit vector in the direction of the $\Lambda$ momentum and 
\begin{equation}
{\vec e}_N = {\vec e}_T\times {\hat p}, \ ~~~~~~ \
{\vec e}_T = \frac{{\vec n} \times {\hat p}}{|{\vec n} \times {\hat p}|}, \ ~~~~~~ \ 
{\hat r} = {\vec e}_T\times {\vec n}.
\end{equation}
Note that the first term of the expression (\ref{dm11}) corresponds to the cases 
where
either $\lambda'_1 = \lambda_1$,  $\lambda'_2 = \lambda_2$ (if $\lambda' = \lambda$) 
or 
$\lambda'_1 \neq \lambda_1$,  $\lambda'_2 \neq \lambda_2$ (if $\lambda' = -\lambda$). 
Conversely the second term corresponds to the cases where either $\lambda'_1 = 
\lambda_1$,  $\lambda'_2 \neq \lambda_2$ (if $\lambda' = \lambda$) or $\lambda'_1 
\neq \lambda_1$,  $\lambda'_2 = \lambda_2$ (if $\lambda' = -\lambda$). We have to 
take into account such combinations in calculating average values of operators, 
according to eq. (\ref{oper}). In the case of observables connected to $V$ it is 
convenient to re-express the SDM as 
\begin {eqnarray}
{\rho}^f_{\mu \mu' \Lambda' \Lambda''} \ &=& \ \frac{1}{4\pi} \Big[A_{\mu 
+\Lambda',\mu} A^*_{\mu'+\Lambda',\mu'}(1+4\Lambda'P^{\Lambda_b}_1) 
\delta_{\Lambda',\Lambda''} \nonumber \\
&+& 2A_{\mu+\Lambda',\mu} A^*_{\mu'-\Lambda',\mu'} (P^{\Lambda_b}_2 + 2i \Lambda' 
P^{\Lambda_b}_3)\delta_{\Lambda',-\Lambda''}\Big], \label{dm22}
\end{eqnarray}
with the constraint $|\mu+\Lambda'|$ = $|\mu'\pm\Lambda'|$ = 1/2.

\subsection{Angular Distribution}

The angular distribution of the decay products, $W(\theta,\phi)$, can be deduced from 
the SDM, according to the formulae
\begin{equation}
W(\theta,\phi) \ = \   Tr \rho^f.
\end{equation}
Taking into account eq. (\ref{dm11}) or (\ref{dm22}), we get
\begin{eqnarray}
W(\theta,\phi) \ = \ {\frac{1}{4\pi}}(G_W \ + \  {\Delta}G_W  \  P^{\Lambda_b}_1), 
\label{dst}
\end{eqnarray}
\noindent
with
\begin{eqnarray}
G_W  \  & = & \  {|A_{1/2, 0}|}^2 + {|A_{-1/2, -1}|}^2 + {|A_{1/2, 1}|}^2 + 
{|A_{-1/2, 0}|}^2, \label{gw}\\
{\Delta}G_W  \ & = &  \ 2\left({|A_{1/2, 0}|}^2 + {|A_{-1/2, -1}|}^2 - {|A_{1/2, 
1}|}^2 - {|A_{-1/2, 0}|}^2\right). \label{dgw}
\end{eqnarray}
We may also obtain the respective projections over the polar and azimuthal angles: 
\begin{eqnarray}
W_p(\theta) & = & {\frac{1}{2}} (G_W  + {\Delta}G_W  P^{\Lambda_b}_z cos \theta),  
\\ 
W_a(\phi)  & = &  {\frac{1}{2\pi}} [G_W  + {\Delta}G_W (P^{\Lambda_b}_x cos \phi + 
P^{\Lambda_b}_y sin \phi )].
\end{eqnarray}
\vskip 0.2cm
\noindent
It is worth noting the crucial role played by the initial polarization of $\Lambda_b$  
in both the polar and azimuthal projections. In particular, the $\phi$-dependence 
disappears if parity is conserved in the production reaction of this resonance.

\subsection{Polarization Vectors}

\noindent
In order to compute the polarization vector of each resonance $R_i$, a special frame 
has to be defined, by means of three mutually orthogonal unit vectors.
For the $\Lambda$ resonance we have
$$  {\hat z^{\prime}} \ = \ {\vec e}_L \ = \ {\hat p},  \ ~~~~ \  {\hat y^{\prime}} \ 
=  \ {\vec e}_T,  \ ~~~~ \  {\hat x^{\prime}} \ = \ {\vec e}_N.  $$
\noindent
The $\Lambda$ polarization vector is decomposed like $ \ \vec {\cal P}^{\Lambda} = 
P_L {\vec e}_L + P_T {\vec e}_T + P_N {\vec e}_N $, where $P_L, \ P_T  \ {\mathrm 
and} \ P_N$ are defined, respectively, as the {\it longitudinal, transverse} and {\it 
normal} component of  $\vec {\cal P}^{\Lambda}$. For the $V$-resonance we have
$ \ \vec {\cal P}^V = P_L \vec{e'}_L + P_T \vec{e'}_T + P_N \vec{e'}_N $, with

$$\vec{e'}_L = -\vec{e}_L, ~~~~~~~~~~ \vec{e'}_T \ = \ -\vec{e}_T, ~~~~~~~~~~ 
\vec{e'}_N \ = \ \vec{e}_N.$$
          
 {\it In these particular frames}\cite{ms} we have, for each resonance,

\begin{eqnarray}
\vec{\cal P}^{R_i}  \ = \   \frac{Tr(\rho^f {\vec s} )}{Tr(\rho^f)},  \ \ \ {\mathrm 
whence} \ \ \  {\vec {\cal P}}^{R_i} \ W(\theta, \phi)\ = \   Tr(\rho^f {\vec s}), 
\label{plr}
\end{eqnarray}
where ${\vec s} \equiv (s_x, s_y, s_z)$ denotes the spin vector operator.
\vskip 0.4cm
\noindent
{\underbar {Polarization Vector of $\Lambda$}
\noindent

We calculate the components of the polarization vector of $\Lambda$ by exploiting eq. 
(\ref{dm11}) of the SDM and eq. (\ref{plr}). The longitudinal component reads  
\begin{eqnarray}
W(\theta,\phi) P^{\Lambda}_L(\theta,\phi) \ = \ {\frac{1}{4\pi}}(G^{\Lambda}_L \ + \  
{\Delta}G^{\Lambda}_L  P^{\Lambda_b}_1), \label{plr0} 
\end{eqnarray}
\noindent
where 
 \begin{eqnarray}
 2G^{\Lambda}_L & = & {|A_{1/2, 0}|}^2 - {|A_{-1/2, -1}|}^2 + {|A_{1/2, 1}|}^2 - 
{|A_{-1/2, 0}|}^2 , \label{gllb}\\
 \Delta G^{\Lambda}_L & = & {|A_{1/2, 0}|}^2 - {|A_{-1/2, -1}|}^2 - {|A_{1/2, 1}|}^2 
+ {|A_{-1/2, 0}|}^2.  \label{dglb}
\end{eqnarray}
As to the transverse component, the previous formulae yield
\begin{eqnarray}
W(\theta,\phi) P^{\Lambda}_T(\theta,\phi) \ = \ {\frac{1}{4\pi}}(G^{\Lambda}_T 
P^{\Lambda_b}_2 \ + \  {\Delta}G^{\Lambda}_T  P^{\Lambda_b}_3), \label{pt} 
\end{eqnarray}
\noindent
where 
 \begin{eqnarray}
 G^{\Lambda}_T & = & -2\Im{\Big( A_{1/2, 0}{A^{\star}_{-1/2, 0}} + A_{1/2,1} 
{A^{\star}_{-1/2, -1}} \Big)}, \label{gtlb}\\
 \Delta G^{\Lambda}_T & = & 2\Re{\Big( A_{1/2, 0}{A^{\star}_{-1/2, 0}} - A_{1/2,1} 
{A^{\star}_{-1/2, -1}} \Big)}. \label{dgtb}  
\end{eqnarray}
Lastly, the normal component yields
\begin{eqnarray}
W(\theta,\phi) P^{\Lambda}_N(\theta,\phi) \ = \ {\frac{1}{4\pi}}(G^{\Lambda}_N 
P^{\Lambda_b}_2 \ + \  {\Delta}G^{\Lambda}_N  P^{\Lambda_b}_3), \label{pn} 
\end{eqnarray}
\noindent
where
\begin{eqnarray} 
G^{\Lambda}_N & = &  2\Re{\Big( A_{1/2, 0}{A^{\star}_{-1/2, 0}} + A_{1/2,1} 
{A^{\star}_{-1/2, -1}} \Big)}, \label{gn}\\
\Delta G^{\Lambda}_N & = & -2\Im{\Big( A_{1/2, 0}{A^{\star}_{-1/2, 0}} - A_{1/2,1} 
{A^{\star}_{-1/2, -1}} \Big)}. \label{dgn}
\end{eqnarray}

\vskip 0.4cm
\noindent
{\underbar {Polarization Vector of $V$}
\noindent

In order to calculate the components of the polarization vector of $V$ we exploit eq. 
(\ref{dm22}) of the SDM and take into account the expression of ${\vec s}$ for spin-1 
particles\cite{bal}. We have
\begin{eqnarray}
W(\theta,\phi) P^V_L(\theta,\phi) \ &=& \ {\frac{1}{4\pi}}(\Delta G^V_L \ + \  G^V_L  
P^{\Lambda_b}_1), \label{pv1}
\\
W(\theta,\phi) P^V_T(\theta,\phi) \ &=& \ {\frac{1}{4\pi}}( G^V_T P^{\Lambda_b}_2\ + 
\  \Delta G^V_T  P^{\Lambda_b}_3), \label{ptv}
\\
W(\theta,\phi) P^V_N(\theta,\phi) \ &=& \ {\frac{1}{4\pi}}(\Delta G^V_T 
P^{\Lambda_b}_2\ - \  G^V_T  P^{\Lambda_b}_3).
\end{eqnarray}
\noindent
Here 
\begin {eqnarray}
G^V_L & = & -2({|A_{-1/2, -1}|}^2 + {|A_{1/2, 1}|}^2),  \\
{\Delta}G^V_L & = & {|A_{1/2, 1}|}^2 - {|A_{-1/2, -1}|}^2,  \\
G^V_T & = & -2{\sqrt{2}} \Im{( A_{1/2,1} A^{\star}_{1/2,0} - A_{-1/2,-1} 
A^{\star}_{-1/2,0} )}, \\
{\Delta}G^V_T & = &  2{\sqrt{2}} \Re{( A_{1/2,1} A^{\star}_{1/2,0} + A_{-1/2,-1} 
A^{\star}_{-1/2,0})}. \label{fnl}
\end{eqnarray}

\vskip 0.4cm
\noindent
{\underbar {Polarization Correlations}
\noindent

Now we define the following four polarization correlations, similar to those 
considered by Chiang and Wolfenstein\cite{cw}:
\begin{eqnarray}
W(\theta,\phi) P_{TT(NN)}(\theta,\phi)  = 
\frac{1}{2}Tr\Big[\rho^f\sigma^{\Lambda}_{y(x)}s^V_{y(x)}\Big],
\\
W(\theta,\phi) P_{TN(NT)}(\theta,\phi)  = 
\frac{1}{2}Tr\Big[\rho^f\sigma^{\Lambda}_{y(x)}s^V_{x(y)}\Big]. \label{doubl}
\end{eqnarray}
These observables are related to the angular correlations of the decay products of 
the $\Lambda$ and $V$ resonance, similar to those considered in refs. 
\cite{va,ddl,dln} and measured in experiments quoted in ref. \cite{dln}. 

Substituting expression (\ref{dm11}) or (\ref{dm22}) into eqs. (\ref{doubl}), we get
\begin {eqnarray}
W(\theta,\phi)  P_{TT}(\theta,\phi) &=&  {\frac{1}{4\pi}}    (G_{TT} \ + \  \Delta 
G_{TT}  P^{\Lambda_b}_1),\\
W(\theta,\phi)  P_{NT}(\theta,\phi) &=& {\frac{1}{4\pi}}    (\Delta G_{TN} \ + \ 
G_{TN}  P^{\Lambda_b}_1), \label{db1} \\
W(\theta,\phi)  P_{TN}(\theta,\phi) &=& {\frac{1}{4\pi}}    (G_{TN} \ + \  \Delta 
G_{TN}  P^{\Lambda_b}_1), \label{db2} \\
W(\theta,\phi)  P_{NN}(\theta,\phi) &=& -{\frac{1}{4\pi}}    ( G_{TT} \ + \ \Delta 
G_{TT}  P^{\Lambda_b}_1),
\end{eqnarray}
with
\begin {eqnarray}
G_{TT} & = & -\frac{1}{\sqrt{2}}\Re{( A_{-1/2,-1} A^{\star}_{1/2,0} + A_{1/2,1} 
A^{\star}_{-1/2,0})}, \label{gtt} \\
\Delta G_{TT} & = & -{\sqrt{2}} \Re{( A_{-1/2,-1} A^{\star}_{1/2,0} - A_{1/2,1} 
A^{\star}_{-1/2,0})}, \label{dgtt}\\
G_{TN} & = & \sqrt{2} \Im{( A_{-1/2,-1} A^{\star}_{1/2,0} + A_{1/2,1} 
A^{\star}_{-1/2,0})}, \label{gtn}\\
\Delta G_{TN} & = & \frac{1}{\sqrt{2}}\Im{( A_{-1/2,-1} A^{\star}_{1/2,0} - A_{1/2,1} 
A^{\star}_{-1/2,0})}. \label{dpol}
\end{eqnarray}

\section{Parametrization of Observables}

In this section we write a model independent parametrization, based on the previous 
formulae, of the angular distribution, of the polarization of $\Lambda$ and of the 
polarization correlations $P_{TT}$ and $P_{TN}$. In particular, we describe such 
observables in terms of a minimum number of independent parameters. The polarization 
components of $V$ can be expressed as functions of such parameters, as is 
straightforward to see from eqs (\ref{pv1}) to (\ref{fnl}).

The formulae of the angular distribution and of the polarization of $\Lambda$ can be 
rewritten as
\begin{eqnarray}
W(\theta,\phi) &=& \frac{1}{4\pi}G_W(1+2P^{\Lambda_b}_1\alpha_W), ~~~~~~~~ \ ~~~~~~~ 
\
~~~~~~~~ \ ~~~~~~~ \ ~~~~~~~~ \ ~~~~~~~ \label{angd}
\\
\vec{{\cal P}}^{\Lambda}(\theta,\phi) &=& \frac{1}{1+2P^{\Lambda_b}_1\alpha_W} 
\left[C_L\vec{e}_L+C_T\vec{e}_T+C_N\vec{e}_N\right], \label{pol1}
\end{eqnarray}
with
\begin{eqnarray}
C_L &=& B_L(1+2P^{\Lambda_b}_1\alpha_L), ~~~~~~ \ ~~~~~~~~ \ ~ C_T \ = \ 
B_T(P^{\Lambda_b}_2+2P^{\Lambda_b}_3\alpha_T), 
\\
 C_N &=& B_N(P^{\Lambda_b}_2+2P^{\Lambda_b}_3\alpha_N)  ~~~~~~~ \ ~~~~~~~~ \ 
~~~~~~~~~~ \ ~~~~~ \ ~~~~~~~~
\end{eqnarray}
and
\begin{eqnarray}
B_L &=& \frac{G^{\Lambda}_L}{G_W}, ~~~~~~~ B_T \ = \ \frac{G^{\Lambda}_T}{G_W}, 
~~~~~~~~~~~ B_N \ = \ \frac{G^{\Lambda}_N}{G_W}, \label{bp}
\\
\alpha_L &=& \frac{\Delta G^{\Lambda}_L}{2G^{\Lambda}_L}, ~~~~~~ \alpha_T \ = \ 
\frac{\Delta G^{\Lambda}_T}{2G^{\Lambda}_T}, ~~~~~~~~~ \alpha_N \ = \ \frac{\Delta 
G^{\Lambda}_N}{2G^{\Lambda}_N}. \label{parm}
\end{eqnarray}
As for the polarization correlations, we have 
\begin{eqnarray}
P_{TT} &=& \frac{1}{1+2P^{\Lambda_b}_1\alpha_W} 
B_{TT}(1+2P^{\Lambda_b}_1\alpha_{TT}), \label{dpol1}
\\
P_{TN} &=& \frac{1}{1+2P^{\Lambda_b}_1\alpha_W} 
B_{TN}(1+2P^{\Lambda_b}_1\alpha_{TN}), \label{dpol2}
\end{eqnarray}
where
\begin{eqnarray}
B_{TT} &=& \frac{G_{TT}}{G_W}, ~~~~~~~ B_{TN} \ = \ \frac{G_{TN}}{G_W}, \label{bpd}
\\
\alpha_{TT} &=& \frac{\Delta G_{TT}}{2G_{TT}}, ~~~~~~ \alpha_{TN} \ = \ \frac{\Delta 
G_{TN}}{2G_{TN}}. \label{dpar}
\end{eqnarray}

The parameters which appear in eqs. (\ref{angd}) to (\ref{dpar}) are not all 
independent of one another, they fulfil the following relations: 
\begin{eqnarray}
B_L^2 (1-\alpha_L)^2 &+& B_{TT}^2 (1-\alpha_{TT})^2 + B_{TN}^2 (1-\alpha_{TN})^2 \ = 
\ (1-\alpha_W)^2, 
\\
B_L^2 (1+\alpha_L)^2 &+& B_{TT}^2 (1+\alpha_{TT})^2 + B_{TN}^2 (1+\alpha_{TN})^2 \ = 
\ (1+\alpha_W)^2,
\\
(\frac{1}{2}\alpha_W-B_L)^2 &+& (B_T-2\alpha_NB_N)^2+(B_N-2\alpha_TB_T)^2 = 
\frac{1}{4}(1-2\alpha_LB_L)^2,
\\
(\frac{1}{2}\alpha_W+B_L)^2 &+& (B_T+2\alpha_NB_N)^2+(B_N+2\alpha_TB_T)^2 = 
\frac{1}{4}(1+2\alpha_LB_L)^2.
\end{eqnarray}
The first two equations allow to express some of the parameters just introduced as 
functions of a more restricted number of other, independent, parameters. We propose 
the following parametrization, similar to previous conventions in hyperon 
decays\cite{ly}:
\begin{eqnarray} 
\alpha_L  \ &=& \ \frac{1-\xi_L}{1+\xi_L}, ~~~~~~~ \alpha_{TT} \ = \ 
\frac{1-\xi_{TT}}{1+\xi_T}, ~~~~~~ \alpha_{TN} \ = \ \frac{1-\xi_{TN}}{1+\xi_{TN}}, 
\\
\xi_L  \ &=& \ \xi_W\frac{cos\psi_-}{cos\psi_+}, ~~~~~~ \ ~~~~~~ \ ~~~~~~~ \ 
~~~~~~~~~ \  \xi_{TT} \ = \ \xi_W\frac{sin\psi_-cos\beta_-}{sin\psi_+cos\beta_+},  
\\
\xi_{TN}  \ &=& \ \xi_W\frac{sin\psi_-sin\beta_-}{sin\psi_+sin\beta_+}, ~~~~~~ \ ~~~~ 
\ ~~~~~~~~ \ \xi_W \ = \ \frac{1-\alpha_W}{1+\alpha_W},  
\\
B_L  \ &=& \ \frac{1\pm\alpha_W}{1\pm\alpha_L} cos{\psi_{\pm}}, ~~~~~~~ \ ~~~~~ \ 
~~~~~~~~ \  B_{TT} \ = \ \frac{1\pm\alpha_W}{1\pm\alpha_{TT}} 
sin\psi_{\pm}cos\beta_{\pm},
\\
B_{TN}  \ &=& \ \frac{1\pm\alpha_W}{1\pm\alpha_{TN}} sin\psi_{\pm}sin\beta_{\pm}, 
~~~~~ \ ~~~~~~~~ \ ~~~~~~~ \ ~~~~~~~ \ ~~~~~~~~~~ \ 
\\
B_T \ &=& \ 1/4(\Gamma_+cos\varphi_+ + \Gamma_-cos\varphi_-), 
~~~~~  B_T\alpha_T \ = \ 1/4(\Gamma_+sin\varphi_+ - \Gamma_-sin\varphi_-), 
\\
B_N \ &=& \ 1/4(\Gamma_+sin\varphi_+ + \Gamma_-sin\varphi_-), 
~~~~~  B_N\alpha_N \ = \ 1/4(\Gamma_+cos\varphi_+ - \Gamma_-cos\varphi_-),
\\
\Gamma \ &=& \ \Big[(1-2\alpha_LB_L)^2-(\alpha_W-2B_L)^2\Big]^{1/2}. \ ~~~~~ \ 
~~~~~~~ \ ~~~~~~~~ \ ~~~~~~ \ ~~~~~~ \
\end{eqnarray}
 
Then the angular distribution, the $\Lambda$ polarization and the polarization 
correlations are expressed as functions of the 10 independent parameters 
$P_1^{\Lambda_b}$, $P_2^{\Lambda_b}$, $P_3^{\Lambda_b}$, $\alpha_W$, $\psi_{\pm}$, 
$\beta_{\pm}$ and $\varphi_{\pm}$.

\section{TRV, CPV and CPT Tests}
Here we illustrate properties of the observables illustrated in the preceding 
sections under discrete transformations and suggest possible tests for violation of 
relative symmetries.
\subsection{T Violations}
The rotationally invariant amplitudes introduced in sect. 2 transform under time 
reversal  (TR) in such a way that\cite{chu}
\begin{equation}
A_{\lambda_1, \lambda_2} A^*_{\lambda'_1, \lambda'_2} \rightarrow  A^*_{\lambda_1, 
\lambda_2} A_{\lambda'_1, \lambda'_2}.
\end{equation}
This follows from the antiunitary character of the TR and from helicity invariance 
under this operation. 
Then eqs. (\ref{pt}), (\ref{gtlb}), (\ref{dgtb}), (\ref{ptv}), (\ref{db1}), 
(\ref{db2}), (\ref{gtn}) and (\ref{dpol}) imply, together with eqs. (\ref{scp}), that 
the transverse polarizations $P_T^{\Lambda}$ and $P_T^V$ and the polarization 
correlations $P_{TN}$ and $P_{NT}$ change sign under TR. Such equations imply, 
together with eq. (\ref{gw}) and the second eqs. (\ref{bp}) and (\ref{bpd}), that 
also the parameters $B_T$ and $B_{TN}$ change sign under the same operation. 
Therefore  nonzero values of such observables are signatures of TRV. These are 
promising for detecting possible effects of NP, according to the considerations of 
refs.\cite{alv,bld,bld2}. Quite analogous properties are shared by two-body decays of 
${\bar \Lambda}_b$. 

In this connection it is worth remembering that also the transverse polarization of 
the muon in $K^+$ decays to $\pi^0 \mu^+ \nu_{\mu}$ and to $\gamma \mu^+ \nu_{\mu}$
has been indicated as a possible signature of TRV\cite{pdg}.

It is important to stress that, in order to get TRV observables, two different 
polarizations are needed, either $\Lambda_b$'s and $\Lambda$'s or $V$'s, or the 
simultaneous measurement of $\Lambda$ and $V$ polarizations. In particular we observe 
that these polarizations are connected to T-odd pseudoscalar triple products. For 
example, we have
\begin {eqnarray}
P_{TN} - P_{NT}  \propto   \langle{\vec s}^V\times{\vec \sigma^{\Lambda}}\cdot {\hat 
p}\rangle, 
\end{eqnarray}
brackets denoting average. Similarly,  by combining $\Lambda_b$'s and $\Lambda$'s 
polarizations, according to formulae (\ref{pt}) and (\ref{scp}), we can perform the 
following triple products:
\begin {eqnarray}
P_3^{\Lambda_b} P_T^{\Lambda} \propto \langle\sigma^{\Lambda_b}_r\rangle {\hat r} 
\times \langle\sigma^{\Lambda}_T\rangle \cdot {\hat p}, ~~~~~~~ \ ~~~~ 
P_2^{\Lambda_b} P_T^{\Lambda} \propto \langle\sigma^{\Lambda_b}_N\langle {\vec e}_N 
\times \langle\sigma^{\Lambda}_T\rangle \cdot {\hat p},
\end{eqnarray} 
where we have set, for the sake of brevity, $\sigma_r$ = ${\vec \sigma}\cdot{\hat r}$ 
and so on. Other authors already proposed T-odd triple 
products\cite{va,wf,alv,bld2,bld}, but those considered here are rid of effects of 
final state interactions\cite{va,wf2}. Moreover, we ascertain {\it a posteriori} that 
some T-odd pseudoscalar\cite{va,ddl,cw,dln} triple products are unequivocally 
connected to TRV.

\subsection{CP Violations}

The CP transformation causes, according to the usual phase conventions\cite{mrr,chu},
\begin{equation}
A_{\lambda_1, \lambda_2}  \rightarrow  -{\bar A}_{-\lambda_1, -\lambda_2},
\end{equation}
where the barred amplitude refers to the ${\bar \Lambda}_b$ decay. Then, taking into 
account eqs. (\ref{bp}),  (\ref{parm}), (\ref{bpd}) and (\ref{dpar}), together with 
the definitions given in subsects. 2.2 and 2.3 of the quantities defined in these 
expressions, we find that the following parameters are useful for detecting possible 
CP violations:
\begin{eqnarray} 
R_W &=& \frac{G_W-{\bar G}_W}{G_W+{\bar G}_W}, ~~~~~~ R_L \ = \ \frac{B_L+{\bar 
B}_L}{B_L-{\bar B}_L},  \label{cp1}
\\
 R_N &=& \frac{B_N-{\bar B}_N}{B_N+{\bar B}_N}, ~~~~~~~ R_{TT} \ = \ 
\frac{B_{TT}-{\bar B}_{TT}}{B_{TT}+{\bar B}_{TT}},
\\
\gamma_W &=& \frac{\alpha_W+{\bar \alpha}_W}{\alpha_W-{\bar \alpha}_W}, ~~~~~~~~ 
\gamma_L \ = \ \frac{\alpha_L+{\bar \alpha}_L}{\alpha_L-{\bar \alpha}_L},
\\
\gamma_T &=& \frac{\alpha_T+{\bar \alpha}_T}{\alpha_T-{\bar \alpha}_T}, ~~~~~~~~~ 
\gamma_N \ = \ \frac{\alpha_N+{\bar \alpha}_N}{\alpha_N-{\bar \alpha}_N},
\\
\gamma_{TT} &=& \frac{\alpha_{TT}+{\bar \alpha}_{TT}}{\alpha_{TT}-{\bar 
\alpha}_{TT}}, ~~~~~~~ \gamma_{TN} \ = \ \frac{\alpha_{TN}+{\bar 
\alpha}_{TN}}{\alpha_{TN}-{\bar \alpha}_{TN}}.
 \label{cp4}
\end{eqnarray}
Any nonzero value of the above defined ratios - defined conformally to the usual 
conventions\cite{chp,tv,hv,va} - would be a signature of CP violation and also, 
possibly, of NP\cite{utf}. The ratios $B_T+{\bar B}_T$ to $B_T-{\bar B}_T$ and 
$B_{TN}+{\bar B}_{TN}$ to $B_{TN}-{\bar B}_{TN}$ have not been considered, since the 
sums are CP-odd and the differences are CPT-odd, therefore both quantities may be, in 
principle, nearly zero. In any case, the sums may be used as  further tests for CP 
violations.   

\subsection{CPT Tests}

The ratios (\ref{cp1}) to (\ref{cp4}) are even under time reversal, therefore they 
can also be suitably employed in tests of the CPT theorem. Moreover it follows from 
the discussion above that also $B_T-{\bar B}_T$ and $B_{TN}-{\bar B}_{TN}$ are good 
parameters for testing the theorem. 

We note that polarization of muons from semileptonic decays of $K^{\pm}$ had been 
proposed by Lee and Wu\cite{lw} as a possible test for CPT violation.

\section{Concluding Remarks}

We conclude this note with some remarks about the method suggested.

A) Our analysis is completely model independent and is also independent of spurious 
effects\cite{alv,bld2,bld,blss,wf} caused by final state interactions\cite{va,wf2}, 
which may flaw, in principle, other kinds of tests proposed\cite{wf,bld2,bs}. In 
particular, we stress that our tests for TRV do not rely on any assumptions. Our 
calculation can be used as an input for calculating the model predictions of the 
observables considered here\cite{aj,aj3}. 

B) It is important to note that the TRV tests based on $\Lambda_b$ polarization are 
similar to those proposed for hyperon decays\cite{ga,chp},
\begin{equation}
\Lambda \to p ~ \pi^-, ~~~~~~ \Sigma \to \Lambda \pi, ~~~~~~ \Xi\to \Lambda \pi.
\end{equation}
However in our case we may also consider the polarization correlations\cite{cw,va}, 
which provide a TRV test independent of the polarization of the parent resonance. 
Decays of the type (\ref{decl}) are very suitable for detecting possible TRV, as 
pointed out also by other authors in studying CP violations\cite{bld2,bld}. 

C) The observables considered in the present letter are very sensitive to NP, since 
they are rid of unpleasant effects of Wilson's coefficients\cite{utf2}. These 
quantities have been considered even more convenient than $B^0-{\bar B}^0$ mixing 
phases\cite{bld2}.

D) Reactions similar to those studied here have been proposed also by other 
authors\cite{leg,les} in a different context, in occasion of LHC forthcoming run. 
Then it appears not unrealistic to suggest to measure also some of the observables 
considered in the present note, that is, the angular distribution and the 
polarization of at least one of the decay products.


\begin{thebibliography}{1}

\bibitem{alv}
T.~M.~Aliev {\it et al.},
\newblock Phys. Lett. B {\bf 542}, 229 (2002)

\bibitem{cgn}
C.~H.~Chen, G.~Q.~Geng and J.~N.~Ng, 
\newblock Phys. Rev. D {\bf 65}, 091502 (2002)

\bibitem{cg}
C.~H.~Chen, G.~Q.~Geng and J.~N.~Ng,
\newblock Nucl. Phys. B (Proc. Suppl.) {\bf 115}, 263 (2003)

\bibitem{bld2}
W.~Bensalem, A.~Datta and D.~London, 
\newblock Phys. Lett. B {\bf 538}, 309 (2002)

\bibitem{bld}
W.~Bensalem, A.~Datta and D.~London, 
\newblock Phys. Rev. D {\bf 66}, 094004 (2002)

\bibitem{blss}
W.~Bensalem, D.~London, N.~Sinha and R.~Sinha,
\newblock Phys. Rev. D {\bf 67}, 034007 (2003)

\bibitem{aj2}
Z.~J.~Ajaltouni, O.~Leitner, P.~Perret, C.~Rimbault and A.~W.~Thomas,
\newblock Eur. Phys. J. C {\bf 29}, 215 (2003)

\bibitem{aj}
Z.~J.~Ajaltouni, E.~Conte and O.~Leitner,
\newblock Phys. Lett. B {\bf 614}, 165 (2005), hep-ph/0412116

\bibitem{aj3}
O.~Leitner and Z.~J.~Ajaltouni,
\newblock Nucl. Phys. B (Proc. Suppl.) {\bf 174}, 169 (2007)

\bibitem{ibl}
M.~Imbeault, S.~Baek and D.~London,
\newblock arXiv:hep-ph/08021175

\bibitem{cpl}
A.~Angelopoulos {\it et al.}, CPLEAR Coll.,
\newblock Phys. Lett. B {\bf 444}, 43 (1998)

\bibitem{chp}
D.~Chang, X.-G.~He and S.~Pakvasa,
\newblock Phys. Rev. Lett. {\bf 74}, 3927 (1995)
  
\bibitem{utf}
M.~Bona {\it et al.}, UTfit coll., 
\newblock arXiv:hep-ph/08030659

\bibitem{ap}
A.~Apostolakis {\it et al.}, CPLEAR Coll.,
\newblock Phys. Lett. B {\bf 456}, 297 (1999)

\bibitem{ak}
L.~Alvarez-Gaum\'e {\it et al.}, 
\newblock Phys. Lett. B {\bf 458}, 347 (1999)

\bibitem{ge}
H.~J.~Gerber,
\newblock Eur. Phys. J. C {\bf 35}, 195 (2004)

\bibitem{fg}
M.~Fidecaro and H.~J.~Gerber,
\newblock Rept. Prog. Phys. {\bf 69}, 1713 (2006)

\bibitem{elm}
J.~Ellis and N.~E.~Mavromatos,
\newblock Phys. Rept. {\bf 320}, 341 (1999)

\bibitem{bes}
J.~S.~Bell and J.~Steinberger,
\newblock Proc. Oxford Int. Conf. Elementary Particles, eds. R.~G.~Moorhouse
et al. (Rutherford Lab., Chilton, England, 1965) p. 195

\bibitem{bs}
I.~I.~Bigi and A.~I.~Sanda,
\newblock Phys. Lett. B {\bf 466}, 33 (1999)

\bibitem{pdg}
Particle~Data~Group,
\newblock Jou. of Phys. G - Nuclear and Particle Physics {\bf 33}, 666 (2006) 

\bibitem{ly}
T.~D.~Lee and C.~N.~Yang,
\newblock Phys. Rev. {\bf 108}, 1645 (1957)

\bibitem{ga}
R.~Gatto,
\newblock Nucl. Phys. {\bf 5}, 183 (1958)

\bibitem{jw}
M.~Jacob and G.~C.~Wick,
\newblock Ann. Phys. {\bf 7}, 404 (1959) 

\bibitem{jk}
J.~D.~Jackson,
\newblock {\it in} High energy physics (Gordon and Breach 
Science Publishers, New York, 1965), p. 327 

\bibitem{bls}
C.~Bourrely, E.~Leader and J.~Soffer,
\newblock {\it Phys. Rept.} {\bf 59}, 95 (1980)

\bibitem{chu}
S.~U.~Chung,
\newblock CERN 71-8 (1971)

\bibitem{ms}
A.~D.~Martin and T.~D.~Spearman,
\newblock "Elementary Particle Theory", North-Holland Publishing Company,
Amsterdam, 1970; p.203 ff.
  
\bibitem{bal}
L.E.~Ballentine,
\newblock "Quantum Mechanics", Prentice-Hall, Inc., Englewood Cliffs,
New Jersey, 1990; p.130 

\bibitem{cw}
C.~W.~Chiang and L.~Wolfenstein, 
\newblock Phys. Rev. D {\bf 61}, 074031 (2000)

\bibitem{va}
G.~Valencia, 
\newblock Phys. Rev. D {\bf 39}, 3339 (1989)

\bibitem{ddl}
A.~S.~Dighe, I.~Dunietz, H.~J.~Lipkin and J.~L.~Rosner, 
\newblock Phys. Lett. B {\bf 369}, 144 (1996)

\bibitem{dln}
A.~Datta {\it et~al.}, 
\newblock Phys. Rev. D {\bf 76}, 034015 (2007)

\bibitem{mrr}
R.~E.~Marshak, Riazuddin and C.~P.~Ryan,
\newblock "Theory of Weak Interactions in Particle Physics",
Wiley Interscience, New York, 1969

\bibitem{lw}
T.~D.~Lee and C.~S.~Wu,
\newblock Ann. Rev. Nucl. Part. Sci. {\bf 16}, 471 (1966) 

\bibitem{tv}
J.~Tandean and G.~Valencia, 
\newblock Phys. Rev. D {\bf 67}, 056001 (2003)

\bibitem{hv}
X.~G.~He and G.~Valencia, 
\newblock Phys. Rev. D {\bf 67}, 056001 (2003)

\bibitem{wf}
L.~Wolfenstein, 
\newblock Phys. Rev. Lett. {\bf 83}, 911 (1999)

\bibitem{wf2}
L.~Wolfenstein, 
\newblock Phys. Rev. D {\bf 43}, 151 (1991)

\bibitem{utf2}
M.~Bona {\it et al.}, LHCb coll., 
\newblock arXiv:hep-ph/07070636

\bibitem{leg}
F.~Legger, LHCb coll.,
\newblock LPHE/2006-002, LHCb/2006-012

\bibitem{les}
F.~Legger and T.~Schietinger, LHCb coll., 
\newblock LPHE/2006-003, LHCb/2006-013

\end{thebibliography}
\end{document}